\documentstyle[twocolumn,prl,aps,epsbox]{revtex}

\newfont{\teneufm}{eufm10 scaled \magstep1}
\newfam\eufmfam
\textfont\eufmfam=\teneufm
\newcommand{\frak}[1]{{\fam\eufmfam\relax#1}}

\begin{document}
\twocolumn[\hsize\textwidth\columnwidth\hsize\csname
@twocolumnfalse\endcsname

\draft
\title{Quantum mutual entropy for Jaynes-Cummings model}
\author{S. Furuichi}
\address{Faculty of Engineering and Sciences, Science University of Tokyo in Yamaguchi, \\ 1-1-1 Daigakudo-ri Onoda city Yamaguchi 756-0884, Japan}
\author{M. Ohya}
\address{Department of Information Sciences, Science University of Tokyo, \\
2641 Yamazaki Noda city Chiba 278-8510, Japan}
\author{H. Suyari}
\address{Department of Information and Image Sciences, Chiba University, \\
1-33 Yayoi-cho Inage-ku Chiba 263-8522, Japan}

\date{\today}

\vspace{1cm}
\maketitle
\vspace{0.5mm}
\begin{abstract}
\quad The dynamics of an atom on the Jaynes-Cummings model has been studied by an atomic inversion, von Neumann entropy and so on.

\quad \, In this letter, we will treat the Jaynes-Cummings model as a problem in non-equilibrium statistical mechanics and apply quantum mutual entropy to study the irreversible dynamics of a state for the atom on this model.
\end{abstract}
\vspace{3.5mm}
\vspace{0.5mm}]

The quantum electrodynamical interaction of a single two-level atom with a single mode of an electromagnetic field is described by the well known Jaynes-Cummings model(JCM)\cite{JC}. 
The JCM is the simplest nontrivial model of two interacting fully quantum systems and has an exact solution. It also brings about some interesting phenomena such as collapses and revivals.
It has been investigated in detail by many researchers from various points of view since \cite{JC}. For details, the reader may refer to the excellent reviews \cite{YE,SK}. The JCM is not only an important problem itself but also gives an excellent example of the interaction between a system and a resevior.
 In this letter, we treat the JCM as a problem in non-equilibrium statistical mechanics and apply quantum mutual entropy\cite{O1} based on von Neumann entropy. For this purpose, we find the quantum mechanical channel\cite{O1} expressing the state change of the atom on the JCM. Our study is an attempt to obtain a new insight of the dynamical change of the state for the atom on the JCM by the quantum mutual entropy.

As is often seen in statistical mechanics, it is one of the most important problems how a system interacting with a reservoir approaches to equilibrium state. Such a problem is often called an open system problem\cite{D}. The open system has been treated by several different methods such as  master equation, von Hove limit and stochastic differential equation. However, in order to uniformly treat many physical transformations as a state change, we have applied the concept of lifting\cite{AO} and the quantum mechanical channel.

We suppose that ${\cal H}_S$ describes an observed system and ${\cal H}_R$ describes a resevior system.
We also suppose that the initial state of ${\cal H}_S$ and ${\cal H}_R$ are represented by $\rho \in {\frak S}_S$ and $\omega \in {\frak S}_R$, respectively. Then, let us consider a continuous mapping ${\cal E}_t^* :{\frak S}_S \to {\frak S}_S \otimes {\frak S}_R$ describing the effect of the interaction between two systems such as;
\begin{equation}
{\cal E}_t^*\rho = U_t \left(\rho \otimes \omega\right)U_t^* \label{l1}
\end{equation}
 where $U_t=\exp\left(-itH/\hbar\right)$ and $H$ is the Hamiltonian of a combined system ${\cal H}_S \otimes {\cal H}_R$. Such a continuous mapping ${\cal E}_t^*$ is called lifting from ${\frak S}_S$ to ${\frak S}_S \otimes {\frak S}_R$.

Our interest is not the dynamical change of the whole system but that of the observed system. In such a case, we have used to compute the {\it reduced density operator}.
That is, if we take the {\it partial trace} with respect to the resevior system ${\cal H}_R$, then we can obtain the final state of the observed system as follows;
\begin{equation}
\Lambda_t^* \rho = tr_{{\cal H}_R} {\cal E}_t^* \rho .\label{c1}
\end{equation}
The mapping $\Lambda_t^* : {\frak S}_S \to {\frak S}_S$ is one of the quantum mechanical channels mapping the initial state $\rho \in {\frak S}_S$ to the final state $\Lambda_t^* \rho \in {\frak S}_S$.  Note that the method to get (\ref{c1}) from (\ref{l1}) by taking a partial trace is widely known as the reduction theory and a lifting expression enables us to provide a general mathematical description for open system\cite{IKO}.

The {\it classical} mutual entropy has been applied to information theory as a quantity of the transmitted information through various physical processes such as a communication process and an irreversible process.
This mutual entropy(information) was defined by using the joint probability. However, the joint probability does not generally exist in quantum system\cite{IU} and the classical mutual entropy is not suited to rigorously analyze the fully quantum system.
In order to formulate the {\it quantum} mutual entropy, a compound state\cite{O1} was introduced in stead of the joint probability.

The quantum mutual entropy was defined in the following manner\cite{O1}. A certain initial state $\rho$ is decomposed as
\begin{eqnarray}
\rho &=& \sum\limits_k {\lambda _kE_k}  \nonumber \\
\lambda _1 & \ge &\lambda _2\ge \cdots \ge \lambda _n\ge \cdots  \nonumber \\
E_i & \bot & E_j \left( {i\ne j} \right) \nonumber
\end{eqnarray}
where $\lambda _k$ is an eigenvalue of $\rho$ and $E_k$ is the associated one dimensional projection. Each $E_k$ can be considered as an elementary event composed of the initial state $\rho$. 
The above decomposition is called (von Neumann) Schatten decomposition\cite{SCHA}. The Schatten decomposition is unique if every $\lambda _k$ is nondegenerated. For a Schatten decomposition, the compound state $\sigma _E$ describing the correlation between an initial state $\rho $ and its final state $\Lambda^* \rho $ is defined by
$$\sigma _E\equiv\sum\limits_k {\lambda _kE_k\otimes \Lambda ^*E_k}. $$
This compound state $\sigma _E$ contains a connection among initial constituents ${E_k}$ and final ones ${\Lambda^* E_k}$ expressing a correlation between $\rho$ and $\Lambda^*\rho$.
Then the quantum mutual entropy, $I\left( {\rho \,;\,\Lambda ^*} \right)$ is defined by
$$I\left( {\rho \,;\,\Lambda ^*} \right)\equiv
\mathop {\sup}\limits_E\left\{ {S\left( {\sigma _E,\;\sigma _0} \right)\;
;\;E=\left\{ {E_k} \right\}} \right\}$$
where $\sigma _0\equiv\rho \otimes \Lambda ^*\rho$ is a trivial compound state and ${S\left( {\sigma _E,\;\sigma _0} \right)}$ is the quantum relative entropy\cite{U} defined by 
$$S\left( {\sigma _E,\;\sigma _0} \right)\equiv
tr\sigma _E\left( {\log \sigma _E-\log \sigma _0} \right). $$
In the above definition of the quantum mutual entropy, we have to take \lq\lq $\sup$" over all Scahtten decompositions when some eigenvalues are degenerated.

The quantum mutual entropy mathematically includes the mutual entropy in classical system.
Moreover, this quantum mutual entropy with the lifting and the quantum mechanical channel mentiond above has been applied to quantum communication theory\cite{IKO,OP}, quantum Markov chains\cite{AOS}, quantum teleportation processes\cite{KOS} and so on. Especially, the irreversibility of a certain system was studied by the quantum mutual entropy in \cite{AOS,O2}.

The quantum mutual entorpy like the classical mutual entropy expresses how much information carried by an initial state $\rho$ is correctly transmitted to a final state $\Lambda^*\rho$ through a quantum mechanical channel. Therefore, since the decrease of the quantum mutual entropy means a loss of initial information, it can be considered a dissipative change of a state of a physical system\cite{O2}. It is a reason why we apply this quantum mutual entropy to investigate the irreversible behavior of the JCM.


The resonant JCM Hamiltonian can be expressed by using rotating-wave approximation in the following form;
\begin{eqnarray}
H&=&H_0+H_1+H_{01} \label{hamil1} \\
H_0&=&\hbar \omega_0 a^*a,\;H_1={1 \over 2}\hbar \omega_0 \sigma _z \label{hamil2}\\
\;H_{01}&=&\hbar g(a\otimes \sigma ^++a^*\otimes \sigma ^-) \label{hamil3}
\end{eqnarray}
where $g$ is a coupling constant, $\sigma^{\pm}$ are the pseudo-spin operators of two-level atom, $\sigma_z$ is the Pauli spin operator, $a$ is the annihilation operator of a photon and $a^*$ is the creation operator of a photon.

In order to derive the quantum mutual entropy, we have to give the quantum mechanical channel for the JCM.
Now, we suppose that the initial state of the atom is a superposition of the upper level and the lower level;
\begin{equation}
\rho = \lambda_0 E_0+\lambda_1 E_1 \in {\frak S}_A  \label{super}
\end{equation}
where, $E_0=\left| 1 \right\rangle \left\langle 1 \right|$, $E_1=\left| 2 \right\rangle \left\langle 2 \right|$, $\lambda_0 + \lambda_1 =1 $.
We also suppose that a field state is a coherent state defined as:
\begin{eqnarray}
\omega &=&\left| \theta  \right\rangle \left\langle \theta  \right|\in {\frak S}_F \nonumber \\
\left| \theta  \right\rangle &=& \exp \left[ {-{1 \over 2}\left| \theta  \right|^2} \right]\sum\limits_l {{{\theta ^l} \over {\sqrt {l! }}}}\left| l \right\rangle 
\end{eqnarray}
Then we set the lifting ${\cal E}^*_t$ and the {\it time-dependent} quantum mechanical channel $\Lambda^*_t$ which describes the time evolution of the atom for the JCM as follows:
\begin{eqnarray}
\Lambda^*_t &:& {\frak S}_A \longrightarrow {\frak S}_A \nonumber \\
{\cal E}^*_t &:& {\frak S}_A \longrightarrow {\frak S}_A \otimes {\frak S}_F \nonumber 
\end{eqnarray}
Then, the final state ${\cal E}^*_t \rho \in {\frak S}_A \otimes {\frak S}_F$ after the interaction between the atom and the photons at the time $t$ is given as follows;
$${\cal E}^*_t\rho = U_t\left(\rho \otimes \omega \right)U^*_t .$$
Moreover the quantum mechanical channel $\Lambda^*_t$ is written by this litfing ${\cal E}^*_t$ as follows;
$$\Lambda ^*_t\rho = tr_{{\cal H}_F} {\cal E}^*_t\rho = tr_{{\cal H}_F} U_t\left( {\rho \otimes \omega} \right)U_t^* .$$
This quantum mechanical channel represents the final state of the atom at the time $t$. 
From the Hamiltonian form in (\ref{hamil1}),(\ref{hamil2}),(\ref{hamil3}),
the following communication relation holds\cite{WM};
\begin{equation}
\left[ {H_0+H_1,H_{01}} \right]=0 .\nonumber
\end{equation}
Therefore the time evolution of the system is determined by the interaction Hamiltonian $H_{01}$;
$$U_t=\exp \left( {-itH_{01}/\hbar } \right) .$$
Take a {\it dressed state}:
$$\left| {\Phi _j^{\left( n \right)}} \right\rangle ={1 \over {\sqrt 2}}\left( {\left| {2\otimes n} \right\rangle +\left( {-1} \right)^j\left| {1\otimes n+1} \right\rangle } \right), \left( {j=0,1} \right)$$
then the following eigen-equation holds:
\begin{equation}
H_{01}\left( \matrix{\left| {\Phi _0^{\left( n \right)}} \right\rangle \hfill\cr\left| {\Phi _1^{\left( n \right)}} \right\rangle \hfill\cr} \right)=\hbar \left( \matrix{\Omega \quad 0\hfill\cr 0\quad -\Omega \hfill\cr} \right)\left( \matrix{\left| {\Phi _0^{\left( n \right)}} \right\rangle \hfill\cr
  \left| {\Phi _1^{\left( n \right)}} \right\rangle \hfill\cr} \right) \label{eigeneq1}
\end{equation}
where $\Omega =g\sqrt {n+1}$ is called the Rabi frequency. From (\ref{eigeneq1}), for $j=0,1$
\begin{equation}
H_{01}\left| {\Phi _j^{\left( n \right)}} \right\rangle =\left( {-1} \right)^j\hbar \Omega \left| {\Phi _j^{\left( n \right)}} \right\rangle \label{eigeneq2}
\end{equation}
Moreover, since for $j=0,1$ we have
$$\left\langle {\Phi _0^{\left( n \right)}} \right|\left. {\Phi _1^{\left( n \right)}} \right\rangle =\left\langle {\Phi _1^{\left( n \right)}} \right|\left. {\Phi _0^{\left( n \right)}} \right\rangle =0,\left\| {\Phi _j^{\left( n \right)}} \right\|=1 .$$
(\ref{eigeneq2}) can be written as;
$$H_{01}=\sum\limits_{n=0}^\infty  H_{01}^{\left( n \right)}=\sum\limits_{n=0}^\infty  {\sum\limits_{j=0}^1 {\left( {-1} \right)^j\hbar \Omega }}\left| {\Phi _j^{\left( n \right)}} \right\rangle \left\langle {\Phi _j^{\left( n \right)}} \right| .$$
Therefore,
\begin{eqnarray}
U_t&=&\exp \left( {-itH_{01}/\hbar } \right) \nonumber \\
&=&\sum\limits_{n=0}^\infty  {\sum\limits_{j=0}^1 E_{n,j}\left| {\Phi _j^{\left( n \right)}} \right\rangle \left\langle {\Phi _j^{\left( n \right)}} \right|} , \label{unitary}
\end{eqnarray}
where $E_{n,j}={\exp \left[ {-it\left( {-1} \right)^j\Omega } \right]}$.

We can compute the transition probability by using this unitary operator.
For example, we now suppose that the initial state of the atom is the upper level. Then the probability $c_n\left(t\right)$ of the atom being in the upper level at the time $t$ is computed as follows;
\begin{eqnarray}
c_n\left(t\right) &=& \vert\langle 2\otimes n \vert U_t \vert 2\otimes n \rangle \vert^2 \nonumber \\
&=& \exp\left[-\vert\theta\vert^2\right]\sum_{n=0}^{\infty}\frac{\vert\theta\vert^{2n}}{n!}\cos^2 \Omega t .\nonumber
\end{eqnarray}
Contrary to this, the probability $s_n\left(t\right)$ of the atom being in the lower level at the time $t$ is also computed as follows;
\begin{eqnarray}
s_n\left(t\right) &=& \vert\langle 1\otimes n+1 \vert U_t \vert 2\otimes n \rangle \vert^2 \nonumber \\
&=& \exp\left[-\vert\theta\vert^2\right]\sum_{n=0}^{\infty}\frac{\vert\theta\vert^{2n}}{n!}\sin^2 \Omega t . \nonumber
\end{eqnarray}

From the unitary operator given in (\ref{unitary}), the final state of the atom of the JCM is given by:
\begin{eqnarray}
\Lambda ^*_t\rho &=& tr_{{\cal H}_F} {\cal E}^*_t\rho = tr_{{\cal H}_F} U_t\left( {\rho \otimes \omega} \right)U_t^* \nonumber \\
&=&\sum\limits_{m,n=0}^\infty  {\sum\limits_{i,j=0}^1 {E_{n,j}E_{m,i}^*\left\langle {\Phi _j^{\left( n \right)}} \right|}}\rho \otimes \omega \left| {\Phi _i^{\left( m \right)}} \right\rangle \nonumber \\
&\times&tr_{{\cal H}_F}\left| {\Phi _j^{\left( n \right)}} \right\rangle \left\langle {\Phi _i^{\left( m \right)}} \right|  \nonumber
\end{eqnarray}
Moreover, from some simple computations, we obtain the quantum mechanical channel such that;
\begin{eqnarray}
\Lambda ^*_t\rho &=&\left( {\lambda_0\tilde c_1\left( t \right)+\lambda_1 \tilde s_0\left( t \right)} \right)E_0 \nonumber \\ 
&+&\left( { \lambda_0\tilde s_1\left( t \right)+\lambda_1 \tilde c_0\left( t \right)} \right)E_1 \label{channel}
\end{eqnarray}
where, $\tilde c_i\left(t\right)=c_{n+i}\left(t\right), \tilde s_i\left(t\right)=s_{n+i}\left(t\right),\, \left(i=0,1\right)$.
This final state is a reduced state for the atom after the interaction with the field. This expression can be seen in the result of another approach by Gea-Banacloche\cite{GB}.

For the computation of the quantum mutual entropy, it is useful to apply the following identity\cite{O1};
$$S\left(\sigma_E,\sigma_0\right)=\sum\limits_k {\lambda _k}S\left( {\Lambda_t^* E_k,\Lambda_t^*\rho } \right) .$$
Since the initial state of the atom given by (\ref{super}) is the non-degenerated Schatten decomposition, the quantum mutual entropy is uniquely given by:
\begin{eqnarray}
I\left( {\rho ;\Lambda_t^*} \right) &=& S\left(\sigma_E,\sigma_0\right)=\sum\limits_{k=0}^1 {\lambda _k}S\left( {\Lambda_t^* E_k,\Lambda_t^*\rho } \right) \nonumber \\
&=&\sum\limits_{k=0}^1 {\lambda _k\sum\limits_{s,t=1}^2 {\left\langle s \right|}}\Lambda_t^*E_k\left| t \right\rangle \log {{\left\langle t \right|\Lambda_t^*E_k\left| s \right\rangle } \over {\left\langle t \right|\Lambda_t^*\rho \left| s \right\rangle }}   \nonumber 
\end{eqnarray}
From the quantum mechanical channel given in (\ref{channel}), the following identities hold:
\begin{eqnarray}
\left\langle 1 \right|\Lambda_t^*E_k\left| 2 \right\rangle &=&\left\langle 2 \right|\Lambda_t^*E_k\left| 1 \right\rangle =0\quad \left( {k=0,1} \right) \nonumber \\
\left\langle 1 \right|\Lambda_t^*E_0\left| 1 \right\rangle &=&\tilde c_1\left( t \right),\left\langle 2 \right|\Lambda_t^*E_0\left| 2 \right\rangle =\tilde s_1\left( t \right) \nonumber \\
\left\langle 1 \right|\Lambda_t^*E_1\left| 1 \right\rangle &=&\tilde s_0\left( t \right),\left\langle 2 \right|\Lambda_t^*E_1\left| 2 \right\rangle =\tilde c_0\left( t \right) \nonumber \\
\left\langle 1 \right|\Lambda_t^*\rho \left| 2 \right\rangle &=&\left\langle 2 \right|\Lambda_t^*\rho \left| 1 \right\rangle =0 \nonumber \\
\left\langle 1 \right|\Lambda_t^*\rho \left| 1 \right\rangle &=& \lambda_0\tilde c_1\left( t \right)+\lambda_1\tilde s_0\left( t \right) \nonumber \\
\left\langle 2 \right|\Lambda_t^*\rho \left| 2 \right\rangle &=&\lambda_0\tilde s_1\left( t \right)+\lambda_1\tilde c_0\left( t \right) \nonumber 
\end{eqnarray}
Therefore, the quantum mechanical entropy can be computed as follows:
\begin{eqnarray}
I\left( {\rho ;\Lambda_t^*} \right)
&=& \lambda_0\tilde c_1\left( t \right)\log {{\tilde c_1\left( t \right)} \over { \lambda_0\tilde c_1\left( t \right)+\lambda_1\tilde s_0\left( t \right)}} \nonumber \\
&+& \lambda_0\tilde s_1\left( t \right)\log {{\tilde s_1\left( t \right)} \over { \lambda_0\tilde s_1\left( t \right)+\lambda_1\tilde c_0\left( t \right)}} \nonumber \\
&+&\lambda_1\tilde s_0\left( t \right)\log {{\tilde s_0\left( t \right)} \over { \lambda_0\tilde c_1\left( t \right)+\lambda_1\tilde s_0\left( t \right)}} \nonumber \\
&+&\lambda_1\tilde c_0\left( t \right)\log {{\tilde c_0\left( t \right)} \over { \lambda_0\tilde s_1\left( t \right)+\lambda_1\tilde c_0\left( t \right)}}  \nonumber
\end{eqnarray}

In FIG.1, the quantum mutual entropy is plotted as a function of time $t$.
What is evident from Fig.1 is that the quantum mutual entropy decreases with the Rabi oscillation. 
Especially, it is obvious that the local maximum point arising in each \lq\lq revival time"\cite{NSE} (i.e. $T_k =k t_r, t_r \cong 2\pi \vert \theta \vert /g , \left(k=0,1,2,\cdot \cdot \cdot \right)$) becomes lower as time goes by.
Therefore, we conclude that the time development of the quantum mutual entropy on the JCM provides a dissipative change of the state of the atom.

\begin{figure}[htbp]
\begin{center}
\epsfile{file=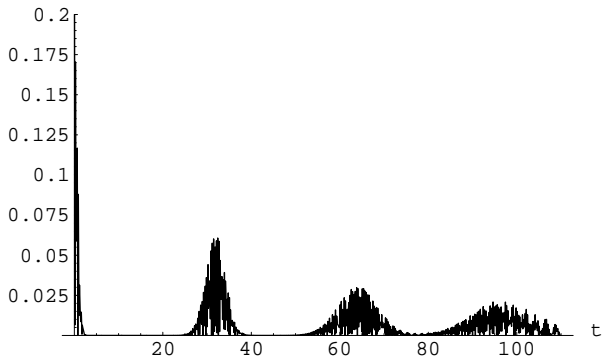,width=8cm,height=5cm}
\caption{The quantum mutual entropy as the mean photon number $\vert \theta \vert^2 =25$, the coupling constant $g=1$, the superposition's parameter of the initial state of the atom $\lambda_0=0.1$, $\lambda_1=0.9$.}
\end{center}
\label{fig1}
\end{figure}

As we have seen, we concretely gave the quantum mechanical channel representing the state change of the atom on the JCM and rigorously derived the quantum mutual entropy. Then, it is shown that the quantum mutual entropy explains the irreversible behavior of the JCM. 
The relation between the decrease of the quantum mutual entropy and the degree of {\it entanglement} can be also studied in the forthcoming paper.


%
%
%
%
%

\end{document}